\def\ts     {\thinspace}
\def\kms    {\ifmmode{{\rm \ts km\ts s}^{-1}}\else{\ts km\ts s$^{-1}$}\fi}
\def\msol   {\ifmmode{{\rm M}_{\odot} }\else{M$_{\odot}$}\fi}
\def\lsol   {\ifmmode{L_{\odot}}\else{$L_{\odot}$}\fi}
\def\lfir   {\ifmmode{L_{\rm FIR}}\else{$L_{\rm FIR}$}\fi}
\def\zsol   {\ifmmode{{\rm Z}_{\odot}}\else{Z$_{\odot}$}\fi}
\def\etal   {{\rm et\ts al.}}
\def\aco    {\ifmmode{{\rm CO}(J\!=\!1\! \to \!0)}\else{{\rm CO}($J$=1$\to$0)}\fi}
\def\bco    {\ifmmode{{\rm CO}(J\!=\!2\! \to \!1)}\else{{\rm CO}($J$=2$\to$1)}\fi}
\def\cco    {\ifmmode{{\rm CO}(J\!=\!3\! \to \!2)}\else{{\rm CO}($J$=3$\to$2)}\fi}
\def\dco    {\ifmmode{{\rm CO}(J\!=\!4\! \to \!3)}\else{{\rm CO}($J$=4$\to$3)}\fi}
\def\eco    {\ifmmode{{\rm CO}(J\!=\!5\! \to \!4)}\else{{\rm CO}($J$=5$\to$4)}\fi}
\def\fco    {\ifmmode{{\rm CO}(J\!=\!6\! \to \!5)}\else{{\rm CO}($J$=6$\to$5)}\fi}
\def\gco    {\ifmmode{{\rm CO}(J\!=\!7\! \to \!6)}\else{{\rm CO}($J$=7$\to$6)}\fi}
\def\hco    {\ifmmode{{\rm CO}(J\!=\!8\! \to \!7)}\else{{\rm CO}($J$=8$\to$7)}\fi}
\def\ico    {\ifmmode{{\rm CO}(J\!=\!9\! \to \!8)}\else{{\rm CO}($J$=9$\to$8)}\fi}
\def\jco    {\ifmmode{{\rm CO}(J\!=\!10\! \to \!9)}\else{{\rm CO}($J$=10$\to$9)}\fi}
\def\kco    {\ifmmode{{\rm CO}(J\!=\!11\! \to \!10)}\else{{\rm CO}($J$=11$\to$10)}\fi}
\def\ci     {\ifmmode{{\rm C}{\rm \small I}}\else{C\ts {\scriptsize I}}\fi}
\def\hi     {\ifmmode{{\rm H}{\rm \small I}}\else{H\ts {\scriptsize I}}\fi}
\def\hh     {\ifmmode{{\rm H}_2}\else{H$_2$}\fi}
\def\cone {\ifmmode{{\rm C}{\rm \small I}(^3\!P_1\!\to^3\!P_0)}
     \else{C\ts {\scriptsize I}{\small$(^3\!P_1\!\to^3\!\!\!P_0)$}}\fi}
\def\ctwo {\ifmmode{{\rm C}{\rm \small I}(^3\!P_2\!\to^3\!P_1)}
     \else{C\ts {\scriptsize I}{\small$(^3\!P_2\!\to^3\!\!\!P_1)$}}\fi}
\def\cij {\ifmmode{{\rm C}{\rm \small I}\,(^3P_i\to^3P_j)}\else{C\ts {\scriptsize I}\,{\small$(^3P_i\to^3P_j)$}}\fi}
\def\cii    {\ifmmode{{\rm C}{\rm \small II}}\else{C\ts {\scriptsize II}}\fi}
\def\tex {\ifmmode{{T}_{\rm ex}}\else{$T_{\rm ex}$}\fi}
\def\tmb {\ifmmode{{T}_{\rm mb}}\else{$T_{\rm mb}$}\fi}
\def\tkin {\ifmmode{{T}_{\rm kin}}\else{$T_{\rm kin}$}\fi}
\def\microns {\ifmmode{\mu{\rm m}}\else{$\mu$m}\fi}
\def\um{\ifmmode{\mu{\rm m}}\else{$\mu$m}\fi}
\def\nhh   {\ifmmode{n({\rm H}_2)}\else{$n$(H$_2$)}\fi}
\def\gradv {\ifmmode{{\rm dv/dr}}\else{dv/dr}\fi}

\documentclass[preprint2]{aastex}
\usepackage{natbib}

\slugcomment{ApJL draft}

\shorttitle{SMM\,J14009+0252}
\shortauthors{Wei\ss, A.\ et al.}

\begin{document}

\title{First redshift determination of an optically/UV faint submillimeter galaxy using CO emission lines}

\author{A.\ Wei\ss\altaffilmark{1}, R.\,J.\ Ivison\altaffilmark{2,3}, D.\ Downes\altaffilmark{4}, F.\ Walter\altaffilmark{5}, M.\ Cirasuolo\altaffilmark{2,3}, K.\,M.\ Menten\altaffilmark{1}}

\altaffiltext{1}{Max-Planck Institut f\"ur Radioastronomy, Auf dem H\"ugel 69, 53121 Bonn, Germany}
\altaffiltext{2}{UK Astronomy Technology Centre, Royal Observatory, Blackford Hill, Edinburgh, EH9 3HJ, UK}
\altaffiltext{3}{Institute for Astronomy, University of Edinburgh, Royal Observatory, Blackford Hill, Edinburgh, EH9 3HJ, UK }
\altaffiltext{4}{Institut de Radio Astronomie Millimetrique, 300 Rue de la Piscine, Domaine Universitaire, 38406 Saint Martin d'H\'eres, France}
\altaffiltext{5}{MPIA, K\"onigstuhl 17, 69117 Heidelberg, Germany}

\begin{abstract}
We report the redshift of a distant, highly obscured submm galaxy
(SMG), based entirely on the detection of its CO line emission.  We
have used the newly commissioned Eight-MIxer Receiver (EMIR) at the
IRAM 30\,m telescope, with its 8\,GHz of instantaneous
dual-polarization bandwidth, to search the 3-mm atmospheric window for
CO emission from SMM\,J14009+0252, a bright SMG detected in the SCUBA Lens
Survey. A detection of the CO(3--2) line in the 3-mm window was confirmed via
observations of CO(5--4) in the 2-mm window. Both lines constrain the
redshift of SMM\,J14009+0252 to $z=2.9344$, with high precision
($\delta z=2\cdot10^{-4}$).  Such observations will become routine in
determining redshifts in the era of the Atacama Large Millimeter/submillimeter Array (ALMA).
\end{abstract}


\section{Introduction} 

Sensitive blank-field mm and submm continuum surveys have
discovered hundreds of dusty, star-forming submm galaxies (SMGs) over the past decade
\citep[e.g.][]{smail97,barger99,borys03,greve04,coppin06}.
Determining their redshift distribution has been much slower, however,
because the large dust content of SMGs means that they often have only weak
(if any) counterparts in the rest-frame ultraviolet and optical,
making spectroscopic redshift determinations extremely difficult
\citep[e.g.][]{smail02,dannerbauer02}. Furthermore, the poor spatial
resolution of mm/submm continuum surveys (typically 11--19$''$) means
that several potential, faint optical/near-IR counterparts exist. 
This requires deep radio or {\it Spitzer} mid-IR images to pinpoint the
most likely counterpart for optical spectroscopic follow-up
observations \citep[e.g.][]{ivison02,ivison04}. The largest SMG
redshift survey published so far was based on radio-identified SMGs
\citep{chapman05}, which may bias the redshift distribution
since radio emission may remain undetected even in the deepest radio
maps for sources at $z>3$.

A promising alternative route to determine the redshift of an SMG is
through observations of CO emission lines at cm or mm
wavelengths. These lines arise from the molecular gas, the fuel for star
formation and can thus be related unambiguously to the submm
continuum source. Therefore, these observations do not require any
additional multi-wavelength identification and circumvent many
of the problems inherent to optical spectroscopy of SMGs.  

\begin{figure*}[t] 
\centering
\includegraphics[width=14.0cm,angle=0]{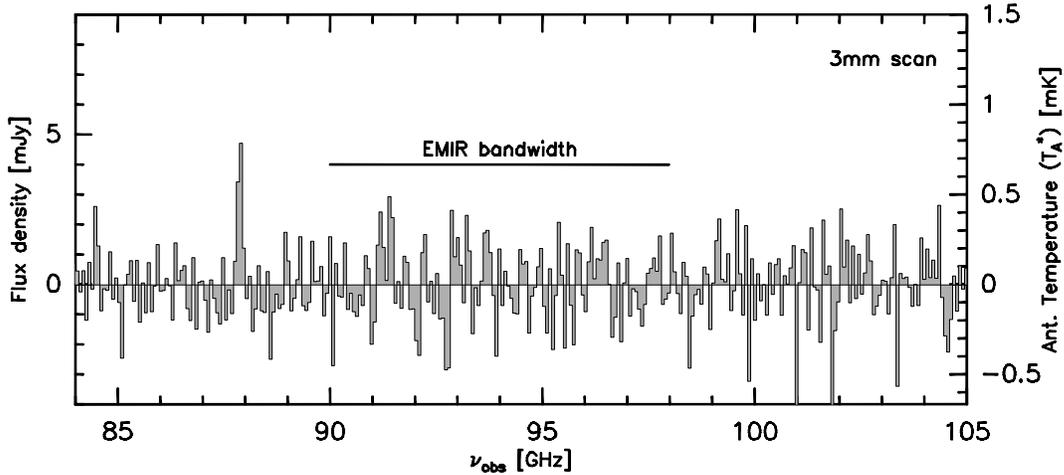} 
\caption{20-GHz-wide spectral scan at a velocity resolution of 200\,\kms\ towards SMM\,J14009+0252 in the 3-mm window. A CO emission
feature is seen at $\sim88$\,GHz (see Fig.\,\ref{spectra} for a presentation of the CO line at higher spectral resolution).}
\label{3mm-scan} 
\end{figure*}

The narrow bandwidth of existing mm receivers, however, placed severe
limitations on this approach as it was too time-consuming to search
blindly for the CO lines in redshift space via multiple frequency tunings. 
In recent years, a lot of effort has been invested in
overcoming this bandwidth limitation at various radio facilities
\citep[e.g.][]{naylor03,erickson07,harris07}. With the commissioning of the
multi-band heterodyne receiver, EMIR, at the IRAM 30\,m telescope,
this situation has greatly improved. EMIR's 8-GHz instantaneous,
dual-polarization bandwidth in the 3-mm band provides the same
spectral coverage as ALMA. Combined with the large
collecting area of the 30\,m telescope this allows for blind searches for
high-redshift CO lines at mm wavelengths.

To demonstrate the capabilities of EMIR as a `redshift machine', we
targeted SMM\,J14009+0252.  This source was discovered by the Submillimeter Common 
User Bolometer Array (SCUBA) in early 1998 and is one of the brightest SMGs discovered 
to date \citep[$S_{850\micron}=15.6$\,mJy,][]{ivison00}. Despite several attempts, and the 
availability of an accurate radio position, no spectroscopic redshift could be determined
-- mainly because of its faintness at near-IR/optical wavelengths. 
In this letter we report the results of our blind search for CO lines in this SMG.

\section{Observing Strategy \label{observations}}

The 3-mm (E090) set-up of EMIR provides 8\,GHz of instantaneous,
dual-polarization bandwidth. The entire accessible frequency range,
$83-117$\,GHz, can be covered with five tunings. This corresponds to
0$<$$z$$<$0.4 for CO(1--0) and 1.0$<$$z$$<$8.7 for the CO lines between ($J=2-1$) and
(7--6), with only a small gap at 1.78$<$$z$$<$1.98.  The gap can be covered
by 2-mm observations, i.e. EMIR is a powerful instrument to search for
high-redshift ($z$$>$1) CO emission.

Observations were made in July 2009 during average summer conditions
($\sim$7\,mm precipitable water vapor). Data were recorded using 16 units of the Wideband Line 
Multiple Autocorrelator (WILMA, 1\,GHz of bandwidth each) to cover 8\,GHz in
both polarizations. WILMA provides a spectral resolution of 2\,MHz
which corresponds to 5-7\,\kms\ for the 3-mm band. The observations
were done in wobbler-switching mode, with a switching frequency of
1\,Hz and an azimuthal wobbler throw of 100$''$.  Pointing was checked
frequently on the nearby quasar J1226+023 and was found to be stable to within
$3''$. Calibration was done every 12\,min using the standard
hot/cold-load absorber. The data were processed with the CLASS software. We
omitted scans with distorted baselines and subtracted only linear
baselines from individual spectra.

\begin{figure*}[t]
\centering
\includegraphics[width=14.0cm,angle=0]{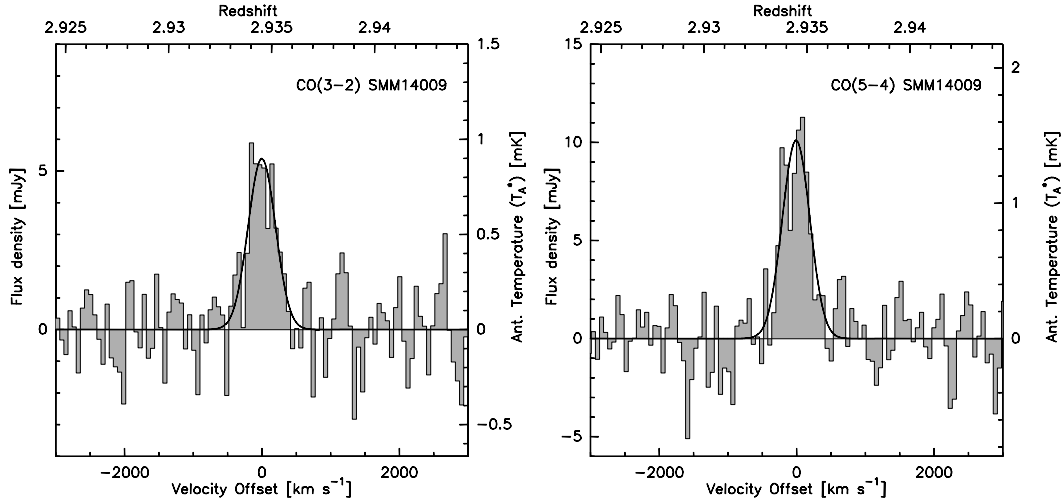}
\caption{Spectra of the CO(3--2) (left) and CO(5--4) (right) lines towards SMM\,J14009+0252. The spectral resolution is 60\,\kms\ for
both lines. See Table\,\ref{colines} for the fit parameters.}
\label{spectra} 
\end{figure*}

We first scanned the full 3-mm tuning range of EMIR with $\sim$2\,hr
of observing for each tuning. The tunings were spaced to provide
500\,MHz overlap. Excellent receiver noise temperatures across the
band (35-45\,K) resulted in typical system temperatures of
$\sim$100\,K. The resulting spectrum had an r.m.s.\ noise level of
0.5\,mK ($\approx 3.5$\,mJy) at a velocity resolution of 200\,\kms\
but did not show clear evidence for CO line emission. We then increased the
integration time for the lower part ($<$105\,GHz) of the 3-mm band
until we reached an average r.m.s.\ noise level of 0.2\,mK (1.2\,mJy).
The resulting spectrum, shown in Fig.\,\ref{3mm-scan}, shows a line at
$\sim$88\,GHz.

At this stage the source redshift was still not determined as it
was not clear which CO transition was detected in the 3-mm scan. We
therefore used the dual-frequency 3-/2-mm (E090/E150) set-up of EMIR to
search for a second CO transition in the 2-mm band and to increase the
signal-to-noise ratio of the 3-mm line. In this configuration, each
frequency band has an instantaneous, dual-polarization bandwidth of
4\,GHz.  The 2-mm mixers were tuned to 146.5\,GHz, under the
assumption that the 3-mm line was the CO(3--2) transition at $z=2.93$. 
At this frequency the receiver noise temperature was $\sim$30\,K, yielding a
system temperature of $\sim$120\,K. SMM\,J14009+0252 was observed in
the dual-frequency set-up for $\sim$5\,hr and we clearly detected a
second line in the 2-mm band (see Fig.\,\ref{spectra}). 
Additional 2-mm data were taken in an attempt to observe a third CO line in the 1-mm band 
(E150/E230 configuration). Given the relatively poor observing conditions, 
the 1-mm data did not yield a meaningful limit.

The beam sizes/antenna gains for the line frequencies at 3 and 2\,mm
are 28$''$/6.0\,Jy\,K$^{-1}$ and 15$''$/6.5\,Jy\,K$^{-1}$,
respectively.  We estimate the flux density scale to be accurate to
$\pm$10--15\%. 

\section{Results} 

The final 3- and 2-mm spectra are shown at a velocity resolution of
60\,\kms\ in Fig.~\ref{spectra}.  The r.m.s.\ noise level ($T_{\rm
A}^*$) for both spectra is 160\,$\mu$K
(1.0\,mJy) and 180\,$\mu$K (1.3\,mJy) at 3 and 2\,mm,
respectively. Both lines are detected at high significance (9 and
12\,$\sigma$ for the integrated intensities). The line profiles for
both lines are very similar and well described by a single Gaussian
with a FWHM of 470\,\kms. The parameters derived from Gaussian fits to
both line profiles are given in Table~\ref{colines}. The frequencies
unambiguously identify the lines as CO(3--2) and CO(5--4) (see our discussion below). 
Combining the centroids of both lines, we derive a variance-weighted mean redshift for
SMM\,J14009+0252 of $z=2.9344 \pm 2\cdot10^{-4}$.

\begin{deluxetable}{l c c c c c c c c}
\tablecaption{CO line parameters for SMM\,J14009+0252.\label{colines}}
\startdata
\tableline
 Line & $\nu_{\rm obs}$ & $z_{\rm CO}$ &$S_{\nu}$ & $\Delta V$ & $I$ & $L'\,\,\,^{a,b}$ & $L\,^{a,b}$\\
            & [GHz]          &       &[mJy]     & [\kms]     & [Jy\kms] & [10$^{10}$\,K\,\kms pc$^2$] & [10$^{7}\,\lsol$]\\

\tableline
CO(3--2) &  \multicolumn{1}{r}{87.888(8)} &2.93450(35) & \multicolumn{1}{r}{$5.4\pm0.9$} & $470\pm60$  & $2.7\pm0.3$ & $7.9\pm0.9 $ & $10.4\pm1.2$\\
CO(5--4) &  \multicolumn{1}{r}{146.469(9)} &2.93438(26) & \multicolumn{1}{r}{$10.2\pm1.3$}& $472\pm45$  & $5.1\pm0.4$ & $5.3\pm0.4 $ & $32.7\pm2.7$\\
\tableline
\enddata
\tablenotetext{a}{corrected for a lens magnification of $m=1.5$ \citep{ivison00}}
\tablenotetext{b}{Adopted luminosity distance: 25.16\,Gpc; angular size distance: 1.625\,Gpc; linear scale: 7.879\,kpc$ /''$ (for $H_0=71\kms$ Mpc$^{-1}$,$\Omega_\Lambda=0.73$ and $\Omega_M=0.27$ \citep{spergel03})}
\end{deluxetable}

\section{Discussion}

At first glance the observed frequencies can not only be
interpreted as CO(3--2) and CO(5--4) at $z=2.93$ but also as CO(6--5)
and CO(10--9) at $z=6.88$ or even CO(9--8) and CO(15--14) at
$z=10.80$. The CO ladder, however, is not equidistant in frequency
which results in small, but significant differences for the frequency
separation of the line-pairs as a function of rotational quantum
number. The frequency separation is 58.577, 58.532 and 58.458\,GHz for
the CO line-pairs at redshifts 2.93, 6.88 and 10.80,
respectively. Our observations yield $\delta\nu=58.581\pm0.017$\,GHz
which identifies the lines as CO(3--2) and CO(5--4) at $z=2.93$.
Our redshift confirms earlier photometric redshift estimates by 
\citet[][$z>2.8$ based on $S_{450}/S_{850}$ and $3<z<5$ based on the whole spectral energy 
distribution (SED)]{ivison00},
\citet[][$z\sim3.5$ based on the dust SED]{yun02} and more recently by \citet[][$z=2.8-3$ 
based on optical/IR photometry]{hempel08}.

With the precise redshift and the observed CO line luminosities in
hand we can estimate the molecular gas content of SMM\,J14009+0252.
The observed CO(5--4) to CO(3--2) line ratio (0.7) implies that
the CO emission is sub-thermally excited, at least for the CO(5--4)
line. This line ratio is identical to that observed for
SMM\,J16359+6612 \citep{weiss05} and we employ the large velocity gradient models
discussed in that paper to estimate a CO(1--0) line luminosity of 
$L'\approx8.2\cdot10^{10}$\,K\,\kms\,pc$^2$.  This translates into a
molecular gas mass of $M_{\hh}\approx6.5\cdot10^{10}\,\msol$ using a
standard ULIRG conversion factor of 0.8\,\msol\,(K\,\kms\,pc$^2$)$^{-1}$ \citep{downes98}. 
These numbers take the lens magnification of $m=1.5$ due to the foreground cluster, Abell
1835 at $z=0.25$, into account \citep{ivison00}.

\begin{figure}[t] 
\centering
\includegraphics[width=7.5cm,angle=0]{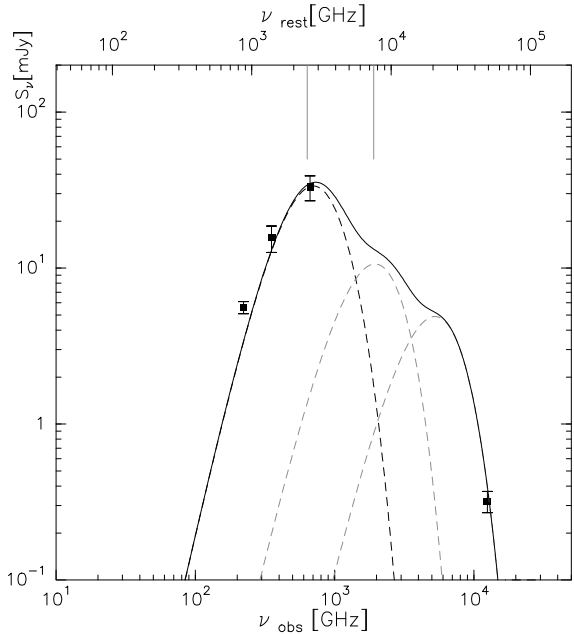} 
\caption{Dust SED towards SMM\,J14009+0252. The black and the two grey dotted lines show a 40, 75 and 200\,K 
dust component, respectively. The far-IR luminosity of this model is $3.8\cdot10^{12}\,\lsol$. 
The solid grey lines at the top indicate the rest-frame 40-120\,\micron\ integration limits used to compute the far-IR luminosity.}
\label{dustsed} 
\end{figure}

The large molecular gas mass is in line with estimates based on the
dust continuum measurements. The 1350-, 850- and 450-\micron\
observations \citep[see][for a compilation of the observed flux
densities]{ivison00} can be described by a dust temperature
of $\sim40$\,K (similar to the kinetic temperature of the CO model) and a
gas mass of $M_{\hh}\approx8\cdot10^{10}\,\msol$ using the dust model in
\citet{weiss07} and a gas-to-dust mass ratio of 100. The implied far-IR
luminosity (integral between 40-120\micron, \citet{helou85}) of this model is $\approx 3\cdot10^{12}\,\lsol$ which corresponds
to a star-formation rate of $\approx 500$\,\msol\,yr$^{-1}$.  These
numbers classify SMM\,J14009+0252 as a ULIRG.

We note, however, that this model underestimates the observed
24-\micron\ flux density and additional warmer dust components are
required to fit the mid-IR data (see Fig.\,\ref{dustsed}). Such a multi-component dust model
predicts $L_{\rm FIR}\approx4-5\cdot10^{12}\,\lsol$, although the lack
of data between 24 and 450\,\micron\ means that the shape of the Wien
tail of the dust SED is not well constrained. In any case the
estimated far-IR luminosity is far ($\sim\,\times\,5$) lower than
estimates based on the radio/far-IR correlation \citep{condon92} which
supports the conclusion of \citet{ivison00} that SMM\,J14009+0252
contains a radio-loud active galactic nucleus.

\acknowledgments 

We thank the IRAM telescope operators for their support during the observations.
IRAM is supported by INSU/CNRS (France), MPG (Germany) and IGN (Spain).

\end{document}